\documentstyle[12pt]{article}

\begin{document}

\input epsf
\renewcommand{\topfraction}{0.9}
\renewcommand{\bottomfraction}{0.9}
\renewcommand{\textfraction}{0.1}
\renewcommand{\floatpagefraction}{0.80}

 
\title{Central Coulomb Effects on Pion Interferometry}
 
\author{D. Hardtke and T.J. Humanic}
 
\maketitle
\begin{center}
Department of Physics, The Ohio State University, Columbus, OH 43210.
\end{center}

\begin{abstract}
Using a simple final-state rescattering model coupled with a simple Coulomb
model, the effects of the central Coulomb potential on pion interferometry
measurements in
158 GeV/nucleon Pb+Pb collisions are explored.  Although the central Coulomb
potential does not introduce correlations between pions, it does prevent an
accurate measurement of the momentum difference.  This
momentum difference smearing effect leads to a reduction in 
the measured correlation radii
and lambda parameters.  These distortions are important in 158 GeV/nucleon Pb+Pb
collisions because of large source sizes and the 
strong central Coulomb potential.  
\end{abstract}
 
\vspace{0.3cm}
 

 
\section{INTRODUCTION}
Pion interferometry is a useful tool for probing the freeze-out conditions in
relativistic heavy-ion collisions, but in order to interpret the measured
correlation radii it is very important to understand any final state
interactions that may distort the measurements.  Recent NA44 
measurements of the transverse mass ($m_T = \sqrt{p_T^2 + m^2}$) dependence 
$\pi^-/\pi^+$ ratio in 158 GeV/nucleon Pb+Pb collisions \cite{coulomb}
show a noticeable enhancement in the $\pi^-/\pi^+$ ratio at low $m_t$.  This
enhancement has been interpreted as evidence for a strong Coulomb interaction
between the pion emitting source and the outgoing pions \cite{coulomb,barz}.  
The strong central Coulomb potential is created by the primary protons in the
projectile and target nuclei.

This work explores the influence of this central Coulomb potential on
two-pion Bose-Einstein correlation measurements.  The single particle data
\cite{coulomb} suggest that central 
Coulomb effects will be important only for low-$p_T$ pion correlation
measurements, but these measurements have been shown to be most sensitive to
the lifetime of the pion emitting source \cite{fields}.  It is thus important
to understand the central Coulomb effects in order to extract useful
information from pion interferometry measurements.  

Although the Au+Au system
at 10.8 GeV/c is not studied explicitly in this work, the qualitative
conclusions should be applicable at these energies since a similar enhancement
in the $\pi^-/\pi^+$ ratio is seen at these energies\cite{ahle,barrette2}.

\section{Rescattering+Coulomb Model}

As input to these calculations, 158 GeV/nucleon Pb+Pb events from a simple
rescattering model were used.
A brief summary of the rescattering calculation used is given below. More
information about this method can be found elsewhere \cite{human94,hum96,hum97}.
Rescattering is
simulated using a Monte Carlo cascade calculation which assumes strong
binary collisions between hadrons. Besides more common hadrons such as pions,
kaons, and nucleons, the calculation also includes the $\rho, \omega, \eta,
\eta^{\prime}, \phi, \Delta$, and $K^{*}$ resonances. 
Resonances can be present at
hadronization and also can be produced as a result of rescattering.
Relativistic kinematics is used throughout. Isospin-averaged scattering
cross sections are taken from Prakash et al \cite{pra93}.

The HBT source sizes extracted from this
model are  in reasonable agreement with the measured source sizes form S+Pb and
S+Ag collisions and preliminary Pb+Pb data \cite{hum96,hum97}.  
The rescattering process generates strong position-momentum
correlations, leading to a $p_T$ dependence of the extracted radius parameters
that matches the experimental observations for the S induced collisions. The
position-momentum correlations produced in the rescattering stage are
important for the Coulomb calculations. 

This model uses isospin averaged cross-sections.  Thus, for example, 
all $\pi^0$, $\pi^+$ and $\pi^-$'s are treated as generic $\pi$'s during the
rescattering stage.  Coulomb interactions are assumed to not be
important during the strong rescattering stage. The effects of the Coulomb
interaction are added after freeze-out using an analytic correction to the
particle momenta.    

The model produces a list of particle freeze-out positions and
momenta.  Coulomb interactions are added by assuming that these particles are
produced in a classical Coulomb potential\cite{baym}. The average 
classical Coulomb
potential is
\begin{equation}
V_s = \frac{Z_{eff}e^2}{r_a},
\label{pot} 
\end{equation}
where $Z_{eff}$ is the effective charge and 
$r_a$ is the average freeze-out radius.  Of course, in a realistic model, each
particle is produced in a different Coulomb potential that depends on
freeze-out time and position.  Using the average Coulomb potential, however, is
computationally much simpler and should not change the qualitative conclusions.

Using simple energy and momentum conservation, the momentum of the particle at
freeze-out can be related to the momentum of 
the particle at the detector using
\begin{equation}
E(p) = E(p_i) + V_s,
\label{cons}
\end{equation}     
where $p$ is the measured momentum of the particle, $p_i$ is the momentum of
the particle at freeze-out, $E(p)=(p^2 + m^2)^{1/2}$, and 
$V_s$ is the effective average Coulomb potential.  
Note that $V_s$ is
positive for positive particles and negative for negative particles.  
Using this equation, the
momentum shift for each particle is calculated:
\begin{equation}
\Delta p = p - p_i.
\label{delp}
\end{equation}
Up to this point the calculation is identical to \cite{baym}.  In
\cite{baym,PBM}, 
the $\pi^+/\pi^-$ ratio at mid-rapidity is calculated to be,
\begin{equation}
\frac{N(\pi^+)}{N(\pi^-)} = \frac{\sqrt{p_{t}^{2} - 2 E(p_t)|V_s| +
V_{s}^2}\;(E(p_t)-|V_s|)} {\sqrt{p_{t}^{2} + 2 E(p_t)|V_s| +
V_{s}^2}\;(E(p_t)+|V_s|)}
\label{sing}
\end{equation} 
The effect of the central Coulomb potential on the measured
momentum difference is calculated to be,
\begin{equation}
Q \approx Q_i (1 + \frac{V_s}{p_i}), 
\end{equation}
where $Q_i$ is the momentum difference at freeze-out and $Q$ is the measured
momentum difference.  This formula predicts that negatively charged particle
pairs will give a measured source size that is larger than the actual source size and
that positively charged pairs will give a measured 
source size that is smaller than
the actual source size.

The simple analytical calculation \cite{baym} does not include any directional
dependence.  Correlation functions are typically measure experimentally in 
three dimensions and parameterized by \cite{NA44dir}
\begin{equation}
C_2(\vec{Q},\vec{k}) = 1 + \lambda exp (-R_{t_{o}}^2(\vec{k}) Q_{t_{o}}^2
-R_{t_{s}}^2(\vec{k}) Q_{t_{s}}^2 -R_{l}^2(\vec{k}) Q_{l}^2),
\end{equation}
where $\vec{k}=\vec{p}_1+\vec{p}_2$ 
is the total momentum of the pair, 
$Q_l$ is the momentum difference in the beam direction, $Q_{t_{o}}$ is the
momentum difference in the direction transverse to beam and parallel to the
total transverse momentum of the pair, and $Q_{t_{s}}$ 
is the momentum difference in the
direction transverse to beam and perpendicular to the total transverse momentum
of the pair.  In this work, the directional dependence is 
explicitly considered. 
$\Delta p$ is calculated analytically using equation (\ref{delp}), 
and then the directional
dependence is added using
\begin{equation}
\vec{p} = \vec{p}_i + \Delta p\; \hat{x},
\label{dir}
\end{equation}
where $\hat{x}$ is the unit vector in the direction of the freeze-out position
$\vec{x}$.  This approach assumes that the central Coulomb charge is spherically
symmetric.

The procedure used in these calculations is the following.  Events were
generated using the simple rescattering model.  This produces a list of pion
freeze-out momenta and positions.  The Coulomb potential is added using
equations
(\ref{cons}),(\ref{delp}), and (\ref{dir}).  The correlation functions are then 
calculated using the
procedure described in \cite{human94,hum96}.  
All correlation functions are evaluated
in the LCMS frame where $p_{z_{1}}+p_{z_{2}}=0$. 
The only free parameter in the
calculation is magnitude of the average Coulomb potential, $V_s$.

\section{Results and Discussion}

The average Coulomb potential $V_s$ is found by fitting equation 
(\ref{sing}) to
the NA44 data \cite{coulomb}.  The value of $V_s$ extracted from the single
particle data is  $\approx$5 MeV. 
This is actually a lower limit on the Coulomb potential since equation
(\ref{sing}) is the mid-rapidity limit ($y=2.9$) and the NA44 data was
measured over the rapidity range $3.1<y<4.1$.  This value of $V_s$ is still
large enough to have effects on the measured pion correlation functions.  

In order to facilitate comparison to future experimental data, 
the pion correlation functions
are calculated for the low $p_T$ NA44 acceptance \cite{NA44dir}.  The rapidity
and $p_T$ range are $0<p_T<400$MeV/c and $3.1<y<4.1$, respectively, 
and the mean $p_T$ is
approximately 150 MeV/c.  
The correlation
function is calculated with three values of $V_s$: $V_s = 0$ (i.e., no Coulomb
effects), $V_s = 5$ MeV (i.e., $\pi^+\pi^+$), and $V_s = -5$ MeV 
(i.e., $\pi^-\pi^-$). 

Figure 1 compares the momentum difference 
shifts in $Q_{t_{s}}$, $Q_{t_{o}}$ and $Q_{l}$
for the positive pion pairs ($V_s = 5$ MeV) and the 
negative pion pairs ($V_s = -5$ MeV).  These momentum difference shifts are then
characterized by a mean momentum shift and an RMS deviation of the momentum
shift, and the values are listed in Table \ref{qshifts}.  Figure 1 and
Table \ref{qshifts} show that the mean shift in all three components of the
momentum difference is small.  This is what would be expected since pions of
similar momentum should experience roughly the same Coulomb impulse.  The RMS
deviations of the momentum shift, however, are significantly larger than the
mean shifts.  
This means that
although the net momentum difference shift is small, the direction of the
momentum difference vector is changed appreciably 
by the central Coulomb potential. 

\begin{table}
\centering
\caption{The mean and RMS deviations of the momentum difference shifts caused
 by the central Coulomb potential for negative and positive pion pairs.}
 \begin{tabular}{|c|c|c|c|c|} \hline
   & \multicolumn{2}{c|}{$\pi^-\pi^-$} 
                      & \multicolumn{2}{c|}{$\pi^+\pi^+$} \\ \hline
       & mean(MeV) & RMS(MeV) & mean(MeV) & RMS(MeV) \\ \hline
$\Delta Q_{t_{s}}$ & -2.4 & 6.4 & 2.5 & 6.3\\ \hline
$\Delta Q_{t_{o}}$ & -1.6 & 6.0 & 1.7 & 5.9 \\ \hline
$\Delta Q_{l}$     & -1.3 & 5.0 & 1.5 & 4.8 \\ \hline
 \end{tabular}
\label{qshifts}
\end{table}
   
The projections of the correlation functions in the NA44 acceptance 
with and without the central
Coulomb potential are shown in Figure 2, and the extracted fit parameters are
listed in Table \ref{fits}.  The extracted radii are smaller for both the 
positive
and negative Coulomb potentials compared to the extracted radii without the
Coulomb potential, and the reduction is largest for $R_L$.  The
magnitude of the reduction in $R_{L}$ and $\lambda$ 
is largest for the negative pions.  The transverse radius parameters
($R_{t_{s}}$ and $R_{t_{s}}$)  are
affected only slightly.    

\begin{table}
\centering
\caption{The fitted results of the gaussian parameterization of the correlation
functions for pions without the central Coulomb potential, 
with a negative Coulomb potential 
($\pi^-\pi^-$), and with a positive
Coulomb potential ($\pi^+\pi^+$).}
 \begin{tabular}{|c|c|c|c|c|} \hline
$V_s$(MeV) & $R_{t_{s}}$(fm) & $R_{t_{o}}$(fm) & $R_{l}$(fm) 
& $\lambda$ \\ \hline
0.     & $5.55 \pm 0.12$ & $6.37 \pm 0.26$ & $8.13 \pm 0.63$ & $0.63\pm0.03$ \\
\hline
-5.0 ($\pi^-\pi^-$) & $5.53 \pm 0.09$ & $6.17 \pm 0.06$ & $6.73 \pm 0.13$ 
& $0.54\pm0.01$ \\ \hline
5.0 ($\pi^+\pi^+$) & $5.33 \pm 0.10$ & $6.14 \pm 0.16$ & $7.38 \pm 0.22$ 
& $0.62\pm0.02$ \\ \hline
 \end{tabular}
\label{fits}
\end{table}

There are two effects due to the central Coulomb potential that lead to a
distortion of the correlation function.  The central Coulomb potential
introduces a net shift in the momentum difference.   
This net shift was pointed
out in \cite{baym}, \cite{barz96}.  
The net shift in the momentum difference, however, has only
a small effect compared to the momentum difference smearing effect.  The
momentum difference smearing effect, characterized by the RMS of the momentum
difference shift distributions, is the more important of the two effects and
leads to a reduction of the measured radii and lambda parameters.  The effect 
is similar to the well understood effect of
the experimental momentum resolution on the correlation function.  The momentum
difference smearing due to the central Coulomb potential is similar for 
$Q_{t_{s}}$, $Q_{t_{o}}$ and $Q_{l}$,  but, due to the larger radius in the
longitudinal direction, the change in the extracted radius is
largest for $R_{l}$.

In \cite{baym} only the net momentum difference shifts were
considered.  This leads to the  
conclusion that positively charged pairs should
measure a source size smaller than the true source size and negatively charged
pairs should measure a source size larger than the true source size.  The
present
work suggests that both negative and positive pairs measure a source size
smaller than the true source size due to the 
momentum difference smearing
effect.  

In \cite{barz96}, the influence of the nuclear Coulomb charge on the measured
transverse correlation radii is much larger than in the present work and
qualitatively different.    
This is
probably due to the different approach used here.  In \cite{barz96}, the 
Klein-Gordon equation is used to find the wave function for a particle
originating in a Coulomb potential parameterized by
\begin{equation}
U = \pm Z e^2 \Phi(r/\sqrt{2}R_0)/r,
\label{barzeq}
\end{equation} 
where $Z$ is the charge number, $R_0$ is the source radius, and $\Phi$ is the
error function.  In \cite{barz96}, the calculation is for Au+Au collisions at 1
GeV/nucleon.  $R_0$ is estimated to be 4.5 fm and $Z$ is taken to be 160. 
Equation (\ref{barzeq}) reduces to equation (\ref{pot}) 
in the limit $r>R_0$.  In
this work, however, equation (\ref{pot}) is not actually used to calculate the
value of the average Coulomb potential $V_s$, but instead a value of $V_s$
is extracted from experimental data.  This leads to a
smaller value for the average 
Coulomb potential than would be extracted from equation
(\ref{barzeq}).  If equation (\ref{pot}) is evaluated with $Z_eff = 160$ and
$r_a = 4.5$ fm, a value of $V_s \approx 50$ MeV is extracted, which should be
compared to value $V_s = 5$ MeV extracted from the experimental data and used
for these calculations.
In addition, it is assumed in this work that the net
nuclear Coulomb charge is spherically symmetric, whereas \cite{barz96} only
looks at the effect of the 
Coulomb field in the transverse direction. The current model has
in fact been examined for the case in which the Coulomb field is assumed to
have only a transverse component, and in that case it was found that the
transverse radii $R_{t_{o}}$ and $R_{t_{s}}$ are reduced much more strongly
than for the central potential used in the calculations presented here.  Even
in the limit where the Coulomb field is assumed to have only a transverse
component, the current work disagrees qualitatively with \cite{barz96} in that
the Coulomb potential reduces all measured correlation radii whereas
\cite{barz96} predicts that the measured sideward radius $R_{t_{s}}$ should be
larger than the true source size for low-$p_T$ $\pi^-\pi^-$ pair measurements.
  
The problem of the central Coulomb potential on the measured pion correlation
functions was also addressed in Ar + KCl collisions at 1.8 GeV/nucleon
\cite{zajc}.  The approach used, however, was much different than that used
here. In \cite{zajc}, an attempt was made to correct the experimental data for
the central Coulomb distortions. 
The momentum shift of each pion due to the central Coulomb charge was
calculated using a formalism correct in both the classical and quantum
mechanical limit \cite{gyulassy}.  It was found that the central Coulomb effect
on the measured correlation function was small compared to the experimental
error on the measured radii.  
The measured radii in \cite{zajc} were much smaller than the predicted radii in
Pb+Pb collisions at 158 GeV/nucleon, so the momentum difference smearing
effect should be much less important.

\section{Conclusions}

Using a simple rescattering model and a classical Coulomb model, it was shown
that the central Coulomb potential in 158 GeV/nucleon Pb+Pb collisions causes a
reduction in the the measured HBT radii and lambda parameters for charged
pions.  This effect should be considered when comparing theoretical predictions
to experimental data.

\section{ACKNOWLEDGEMENTS}
 
We wish to thank Michael Lisa and Nu Xu for useful
discussions.  This work was supported by NSF grant number PHY-9511850.

\section{FIGURE CAPTIONS}

Figure 1: The shifts in $Q_{t_{s}}$, $Q_{t_{o}}$, and $Q_{l}$ due to the
central Coulomb potential for negative and positive pion pairs.  The average
Coulomb potential $V_s = 5$ MeV.

Figure 2: The projections of the correlation functions (C2) in 
$Q_{t_{s}}$, $Q_{t_{o}}$, and $Q_{l}$ for positive and negative pions (open
triangles) compared with the projections without the central Coulomb potential
(open circles).  The fit functions are also included.  The projections are over
the lowest 20 MeV in the other momentum difference directions.

\vspace{0.3cm}
 

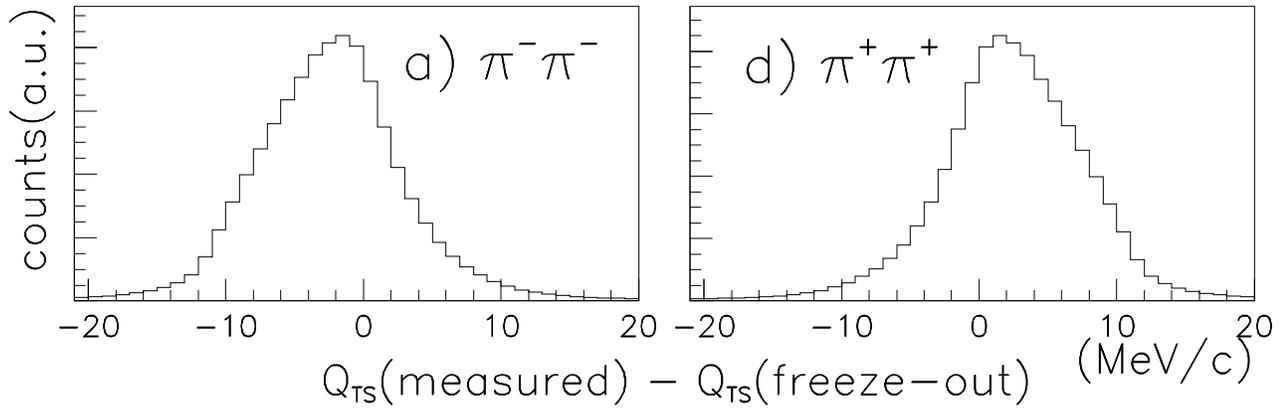
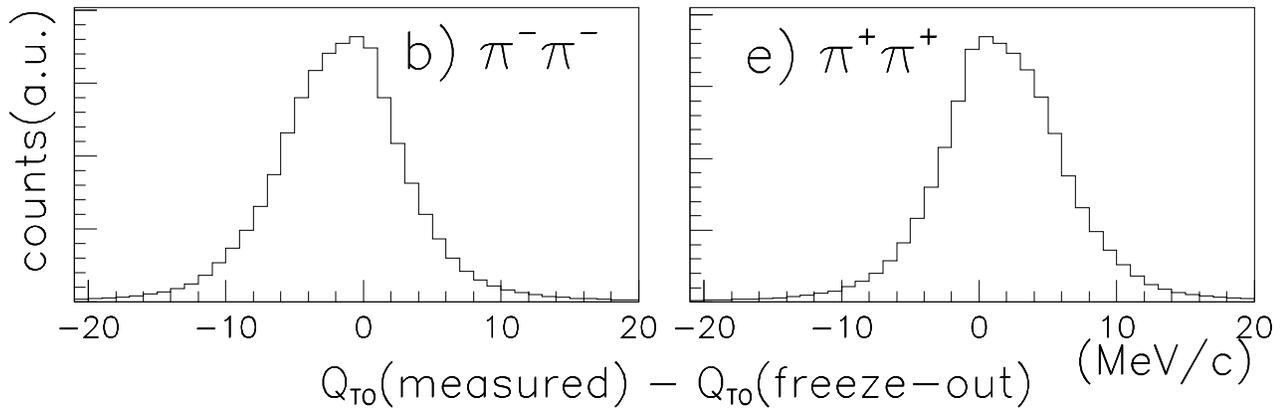
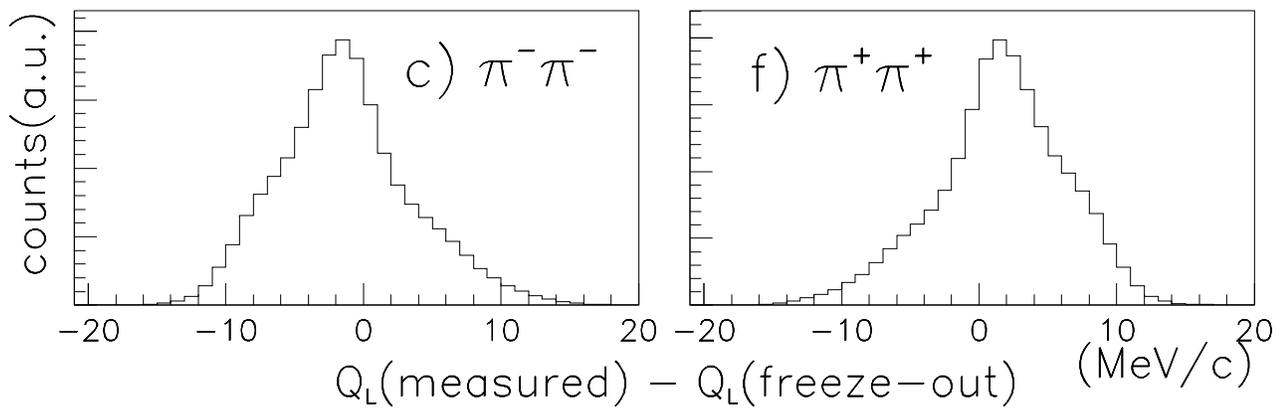
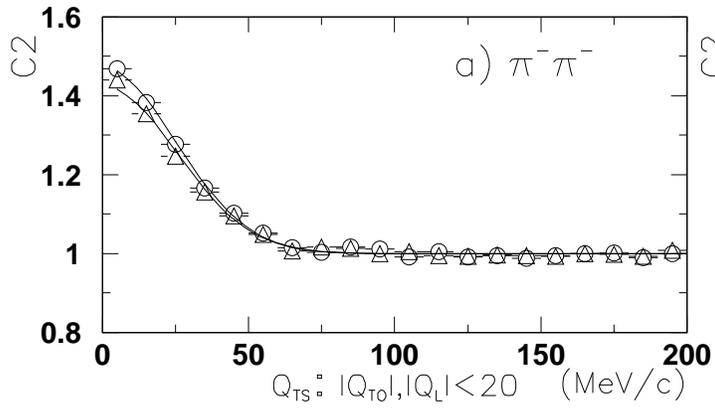
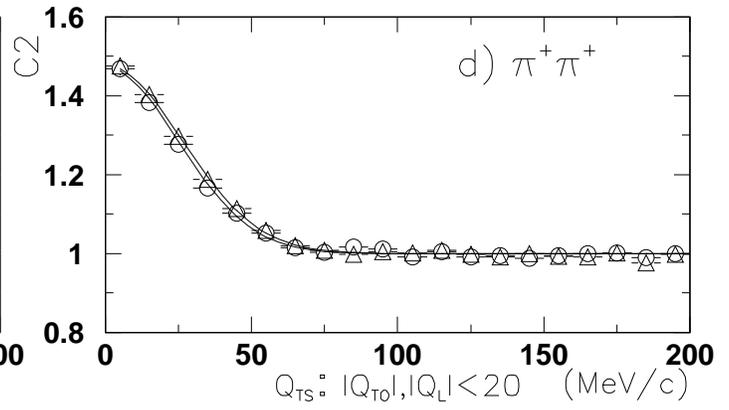
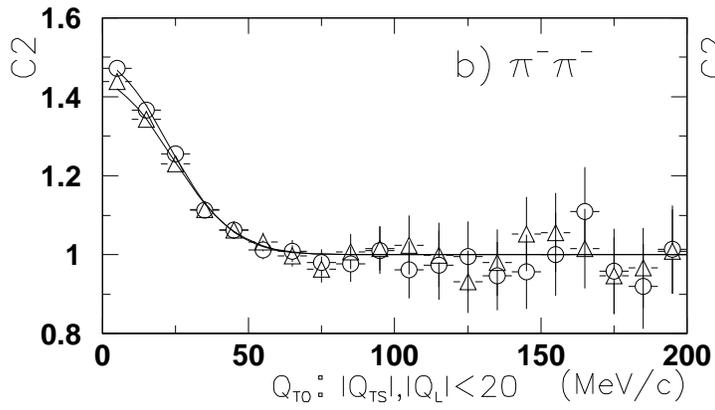
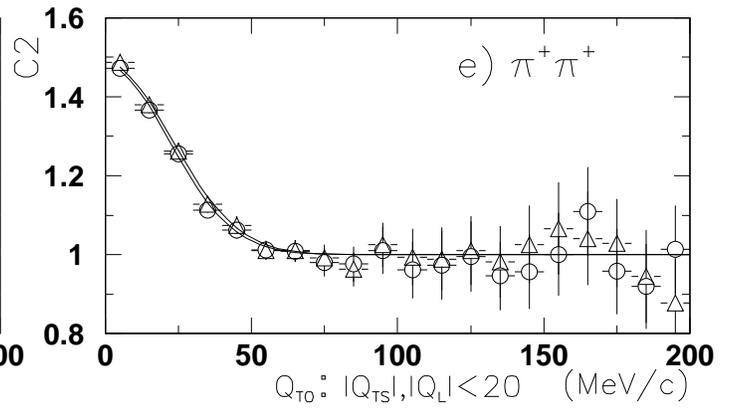
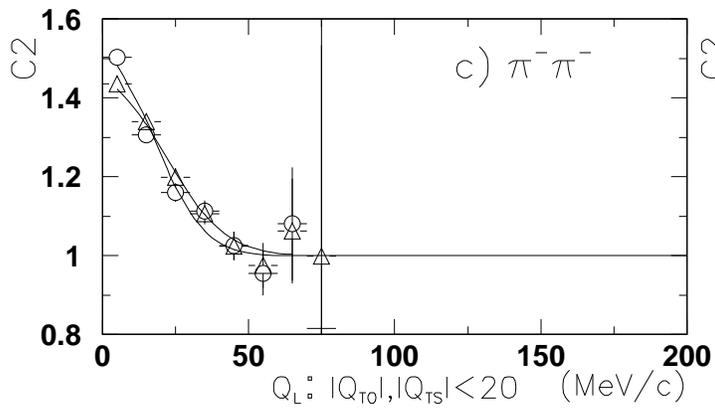
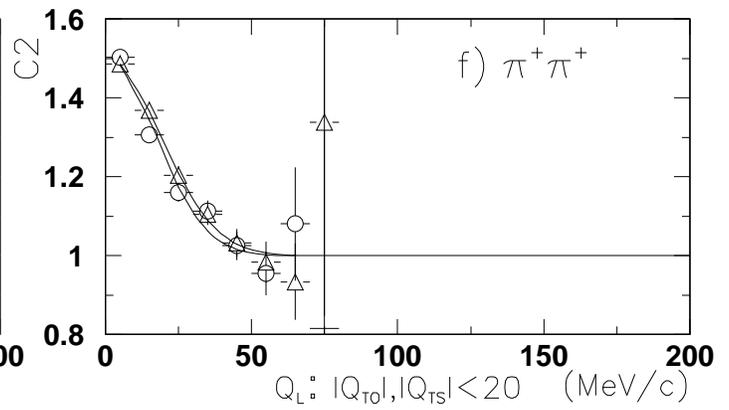
\end{document}